\documentstyle[12pt]{article}
\newcommand{\beq}{\begin{equation}}
\newcommand{\eeq}{\end{equation}}

\begin{document}

\baselineskip=14pt
%\tightlines

\begin {center}

{\bf Trapped Atomic Fermi Gases}\bigskip

M.Ya. Amusia$^{a,b}$,  A.Z. Msezane$^c$, and V.R. Shaginyan$^{c,d}$
\footnote{E--mail: vrshag@thd.pnpi.spb.ru}\\

$^{a\,}$The
Racah Institute of Physics, the Hebrew University, Jerusalem 91904,
Israel;\\ $^{b\,}$Physical-Technical Institute,
194021 St. Petersburg, Russia;\\ $^c\,$CTSPS, Clark Atlanta University,
Atlanta, Georgia 30314, USA;\\
$^{d\,}$Petersburg Nuclear Physics Institute,
Gatchina, 188300, Russia

\end{center}

\begin{abstract}
A many-body system of fermion atoms with a model interaction
characterized by the scattering length $a$ is considered. We treat
both $a$ and the density as parameters assuming that the system can
be created artificially in a trap. If $a$ is negative the system
becomes strongly correlated at densities $\rho \sim |a|^{-3}$,
provided the scattering length is the dominant parameter of the
problem. It means that we consider $|a|$ to be much bigger than the
radius of the interaction or any other relevant parameter of the
system. The density $\rho_{c1}$ at which the compressibility vanishes
is defined by $\rho_{c1}\sim |a|^{-3}$. Thus, a system composed of
fermion atoms with the scattering length $a\rightarrow -\infty$ is
completely unstable at low densities, inevitably collapsing until the
repulsive core stops the density growth. As a result, any Fermi system
possesses the equilibrium density and energy if the bare
particle-particle interaction is sufficiently strong to make $a$
negative and to be the dominant parameter. This behavior can be
realized in a trap. Our results show that a low density neutron matter
can have the equilibrium density.  \end{abstract}

{\it PACS numbers:} 05.30.Fk, 24.10.Cn, 71.10.Ca

{\it Key Words}: trapped Fermi gases; effective interaction\\

Although a theory of Fermi gases is not yet well developed, there
are nevertheless new challenging experimental possibilities
to explore trapped Fermi gases \cite{mj}. These are expected
to stimulate a proper theoretical
description of these Fermi quantum systems. Currently this description of
quantum Fermi gases at their different states, which can be
equilibrium, quasi-equilibrium or far from equilibrium, is usually based
on tedious numerical calculations, particularly when the interaction
between particles, or atoms, of a gas cannot be considered as weak.
This implies that the dimensionless effective coupling constant of the
interaction $p_{F}a\geq 1$. Here $p_{F}$ is the Fermi momentum and is
related to the system's density by
$\rho =p_{F}^{3}/3\pi ^{2}$, while $a$ is the
scattering length. There are a few cases, when it is possible to
treat the properties analytically. The Random Phase Approximation
(RPA) is applicable for a high density electron gas \cite{gb} and the
low density approximation deals with dilute gases \cite{ly}. In both
cases the kinetic energy $T_{k}$ is assumed to be
much bigger than the interaction energy $E_{int}$ of the system.
This permits the application of some kind of a
perturbation theory. In the case of an electron liquid it turns out that
the analytical RPA-like description is also possible not only at very
high but also at medium densities when $T_{k}\sim E_{int}$ \cite{sh,as}.
Similar extension of the range of validity is impossible in the case
of fermion systems at low densities $\rho $ (there the gas
approximation is not applicable if $p_{F}a\geq 1$, or $T_{k}\sim
E_{int}$). If the pair interaction is attractive and sufficiently
strong, the system can have a quasi-equilibrium or equilibrium states
in which $T_{k}\simeq E_{int}$. On the other hand, these states are
separated by special regions at which the incompressibility $K(\rho
)\leq 0$ and the system is completely unstable. An experimental study
performed on such Fermi-systems would be of great importance
presenting new information on the behavior of dilute gases and on the
gas-liquid phase transition. The observation of the existence of
such regions and points at which $K(\rho )=0$ can present a challenging
problem for a theory designed to describe these peculiarities.

In this Letter we address the above mentioned problem and consider
an infinitely extended system composed of Fermi particles, or atoms,
interacting by an artificially constructed potential with the desirable
scattering length. We demonstrate that the consideration can be accomplished
analytically provided that the pair interaction between fermions is
characterized only by the scattering length. That is $|a|$ is much larger
than the other relevant parameters of the interaction, or $a\rightarrow
-\infty$. In this case one can say that the scattering length is the
dominant parameter of the problem under consideration. As it will be
demonstrated below, in such a case the system is located in the
vicinity of the gas-liquid phase transition, transforming it into a
strongly correlated one. Therefore, the problem of calculating its
properties has to be treated for the most part qualitatively. Such an
investigation is of great importance since it can be applied to
fermion systems interacting via potentials with not only infinite,
but also sufficiently large $a$. For instance, the scattering length
$a$ of neutron-neutron interaction is about $-20$ fm, thus greatly
surpassing the radius of the interaction $r_{0}$. Then this
investigation can be viewed as the first step to study trapped Fermi
gases, which are systems composed of Fermi atoms interacting by an
artificially constructed potential with almost any desirable
scattering length, similarly to that done for the trapped
Bose gases, see e.g. \cite{nat}.

Let us start by considering the general properties of a Fermi
system with some attractive two-particle
bare interaction $gV(r)$ of range
$r_{0}$ and $g$ defining the strength of the interaction. We assume
that $gV(r)$ is sufficiently weak to create a two-particle bound
state and that the scattering length $a$ corresponding to this
potential is the dominant parameter, being negative and finite, while
$r_{0}/|a|\ll 1$. Then in the Hartree-Fock approximation the ground
state energy $E_{HF}(\rho)$ is given by the following expression
\begin{equation}
E_{HF}(\rho)=\frac{3p_{F}^{2}}{10M}\rho +t_{HF}\rho ^{2},
\end{equation}
where $M$ is the particle mass, and the parameter $t_{HF}$ , being
negative, is entirely determined by the potential $gV(r)$. For
instance, in the case of a short range $\delta $-type interaction one
has $t_{HF}=-g/4$. Eq. (1) shows that at low densities $E_{HF}>0$ due
to the kinetic energy term, but at sufficiently high densities $\rho
\rightarrow \infty $ the potential energy $t_{HF}\rho ^{2}$
becomes dominating, leading to the collapse
of the system, with $E_{HF}\rightarrow
-\infty $. Keeping in mind that the Hartree-Fock approximation gives
the upper limit to the binding energy $E_{HF}\geq E$, one can
conclude that the system does not have, in this case, an equilibrium
density $\rho_{e}$ and energy $E_{e}$ since $E_{e}\rightarrow -\infty $
when $\rho \rightarrow \infty $ \cite{ll}. Note, that for a given
and finite total number of particles $N$, the HF energy is not going
to infinity and the system collapses into a small volume with the
radius $r_{0}$, with the density $\rho \sim N/r_{0}^{3}$. At the
densities $\rho \rightarrow 0$ we deal with a dilute gas and the
energy $E(\rho )\rightarrow 0,$ remaining positive at these densities
\cite{bk}. Therefore, it must have at least one maximum at the
density $\rho_{m}$ before it becomes negative, on the way to
$E\rightarrow -\infty $. If the potential $gV(r)$ includes a kind of
a ''repulsive core'' at sufficiently short distances, the system has
an equilibrium density and energy, $\rho_{e}$ and $E_{e}$,
respectively, determined by the repulsive core strength and its
radius $r_{c}\sim r_{0}$.
The general features of the function $E(\rho)$
can be qualitatively represented by a simple expression
\begin{equation} E(\rho )=\frac{3p_{F}^{2}}{10M}\rho
+t_{0}\rho^{2}+t_{3}\rho ^{3}. \end{equation} The first term of Eq.
(2) is the kinetic energy $T_{k}$, while the second and the third terms are
related to the interaction energy $E_{int}$ determined by the
potential $gV(r)$. The second term which is proportional to $t_{0}$
gives a qualitative description in the gas limit. The third term
provides the behavior of $E(\rho)$ at higher densities, including
that of the equilibrium density, so that $t_{0}<0$, and $0<t_{3}$.

Now we apply Eq. (2) to demonstrate the most important features of
the system under consideration:

a) when $\rho \rightarrow 0$ the third term on r.h.s. in Eq. (2) can
be omitted. The kinetic energy is relatively very big, $T_{k}\gg
E_{int}$, and $t_{0}\sim a$, with $a<0$ being the scattering
length. In that case we have a dilute Fermi gas with positive
pressure $P$ and incompressibility $K$, the latter being determined
by the equation, see e.g. \cite{lanl}, \begin{equation}
K(\rho)=\rho^{2} \frac{dE^{2}(\rho)}{d\rho^{2}}.
\end{equation}

b) on the way to higher densities, which can be achieved by applying
an external pressure, the system reaches the density $\rho
_{c1}<\rho_{m}$ at which the incompressibility is equal to zero,
$K(\rho_{c1})=0$. Remembering that at the maximum the second
derivative is negative, one can conclude, as it is seen from Eq. (3),
that $K(\rho_{m})<0$. In the range $\rho_{c2}\geq \rho \geq \rho
_{c1}$ the incompressibility is negative, $K<0$, and as a result the
system becomes totally unstable. In fact, in this density range all
calculations of the ground state energy are meaningless since such a
system cannot exist and thus be observed experimentally \cite{lanl};

c) at some point $\rho =\rho_{c2}>\rho_{m}$ the contribution due
to the repulsive core becomes sufficiently strong to prevent the
further collapse of the system. The incompressibility attains $K=0$
at $\rho_{c2}<\rho_{e}$, being positive at the higher densities.
Finally, the system becomes stable at $\rho >\rho_{c2}$, reaching
equilibrium density at $\rho_{e}$ with equilibrium energy equal to
$E_{e}$. Generally speaking, the system is quasi-stable at the
densities $0<\rho <\rho_{c1}$ and $\rho_{c2}<\rho <\rho_{e}$
because, being affected by a sufficiently strong external field, it
occupies its real stable state with the energy $E_{e}$ and density
$\rho_{e}$ \cite{lanl}. For the sake of simplicity, we call these
states stable states. It is obvious that $K(\rho_{e})>0$ being
proportional to the second derivative at the minimum [see Eq. (3)].
It should be kept in mind that in this density domain, $\rho \geq
\rho_{c2}$, the function $E(\rho )$ is determined by the repulsive
part of the potential which makes $t_{3}>0$. As mentioned
above, without this component of $gV(r)$ the system's energy would
decrease, $E_{HF}(\rho)\rightarrow -\infty $, 
with the density growth, $\rho \rightarrow \infty $, 
thus inevitably collapsing.
Indeed, in this case, the potential being pure attractive has no
structure to ensure any stable states at the densities $\rho \geq
\rho_{c1}$. As a result, one can write down a dimensionless
expression for the ground state energy as a function of the only
variable $z=p_{F}a$ \cite{ly,bk}, \begin{equation} \alpha
E(z)=z^{5}(1+\beta (z)), \end{equation} with $\alpha =10\pi
^{2}Ma^{5}$. Note, that the function $\beta (z)$ 
dependents only the variable $z$. In the low density limit,
$|ap_{F}|\ll 1$ and when the interaction has the radius $r_{0}$, Eq.
(4) reads \cite{bk},
\begin{equation}
\alpha E(z)=z^{5}\left[ 1+\frac{10}{9\pi}z+
\frac{4}{21\pi^{2}}(11-2\ln2)
z^{2}+\left(\frac{r_{0}}{a}\right)^{3}
z^{3}\gamma\left(\frac{r_{0}}{a},z\right) +...\right].
\end{equation}
Here the function $\gamma(y,z)$ is of the order of one,
$\gamma(y,z)\sim 1$. It is seen from Eq. (5) that as soon as the
scattering length becomes large enough, $|a|\gg r_{0},$ one can omit
the contribution coming from the function $\gamma $ and neglect all
the term proportional to $(r_{0}/a)^{3}$.  Then Eq. (5) reduces to
Eq. (4). Thus, in the case $|a|\rightarrow \infty $ we can use Eq.
(4) to determine the ground state energy $E$. Eq. (4) is valid up to
the density $\rho_{c1}$ which is a singular point of the function
$\beta (z)$, since beyond this point $K<0$, and the system is
completely unstable. On the other hand, there is no physical reason 
to have another irregular point in the region 
$0< \rho < \rho_{c1}$. Using Eq. (3) 
for the incompressibility and 
Eq. (4) for the energy, one can calculate
the position of the point $z_{c1}$ where $K=0$. Denoting the
corresponding $z$ as $z_{c1}=c_{0}$ , where $c_{0}$ is a
dimensionless number, one is led to the conclusion that $\rho
_{c1}\sim |a|^{-3}$ provided $a$ is sufficiently large to be the only
dominating parameter. As it is seen from Eq. (5), we can expect 
corrections of the order of $(r_0/a)^9$ to this universal behavior.
Thus, the system has only one stable region at
small densities $\rho \leq \rho_{c1}$ which decreases and even vanishes as
soon as $a\rightarrow -\infty$, and the function $\beta $ is in fact 
determined only in the region $|z|\leq |z_{c1}|$ \cite{as1}.

Consider the behavior of the system
when the density approaches $\rho \rightarrow \rho_{c1}\sim
|a|^{-3}$ from the low density side. Normally, points $\rho_{c1}$
and $\rho_{c2}$, overlooked in calculations because of the lack
of self consistency \cite{ksk}, which relates the linear response
function of the system to its incompressibility $K$, \begin{equation}
\chi (q\rightarrow 0,i\omega \rightarrow 0)
=-\left(\frac{d^{2}E}{d\rho ^{2}}
\right)^{-1}.
\end{equation}
These points can give important contributions
to the ground state energy. To see this we express the energy of the 
system in the following form (see e.g. \cite{ksk}), \begin{equation}
E(\rho )=E_{HF}(\rho)-\frac{1}{2}\int \left[ \chi (q,i\omega ,g)-
\chi_0(q,i\omega )\right] v(q)\frac{d{\bf q}\,d\omega\,dg}
{g(2\pi)^{4}}, \end{equation}
where $\chi(q,i\omega,g)$ is the linear
response function on the imaginary axis,
$\chi_0(q,i\omega)$ is the linear response function of 
noninteracting particles, and $v(q)$ is the Fourier
image of $gV(r)$. The integration over $\omega$ goes from
$-\infty$ to $+\infty $, while the integration over the coupling
constant $g$ runs from zero to the real value of the coupling
constant, i.e. to $g=1$. At the point $\rho =\rho_{c1}$ the linear
response function has a pole at the origin of coordinates
$q=0,\,\omega =0$ due to Eq.  (6). At the densities $\rho >\rho
_{c1}$ the function $\chi (q,i\omega,g)$ has poles at finite values of
the momentum $q$ and frequencies $i\omega $.  This prevents the
integration over $i\omega $, making the integral in Eq.  (7)
divergent. Thus, we conclude that it is the contribution of these
poles that reflects the system's instability in the density range
$\rho_{c1}\leq \rho \leq \rho_{c2}$. Note, that violations of Eq.
(6) lead to serious errors in the calculation of the ground state
energy. Equation (7) can be rewritten, explicitly accounting for the
effective interparticle interaction $R(q,i\omega ,g)$,
in the following form \begin{equation} 
E(\rho)=E_{HF}(\rho)-\frac{1}{2}\int \left[ 
\frac{\chi_{0}^2(q,i\omega)R(q,i\omega,g)}
{1-R(q,i\omega,g)\chi_{0}(q,i\omega)}\right]
v(q)\frac{d{\bf q}\,d\omega\,dg}{g(2\pi)^{4}},
\end{equation}
and $\chi$ is given by the following equation \cite{sh,ksk}
\begin{equation}
\chi (q,\omega,g)=\frac{\chi_{0}(q,\omega )}
{1-R(q,\omega,g)\chi_{0}(q,\omega )}.
\end{equation}
It is seen from Eqs. (6) and (9) that the
denominator $(1-R\chi_{0})$ vanishes at
$\rho \rightarrow \rho_{c1}$. As a result, we obtain
\begin{equation}
R(q\to 0,\omega=0,g=1)\propto \frac{1}{\chi_0(q,0)}\sim 
\frac{1}{p_FM+q^2/p_F^2}.\end{equation}
Here $p_F^3\sim \rho_{c1}$. If the scattering length tends
to infinity driving $\rho_{c1}\to 0$, it follows from Eq. (10)
that \beq R(q\to 0,0,1)\propto \frac{1}{q^2},\eeq
and we can conclude that the effective interaction $R$
behaves as a gravitational-like field, leading
the system to collapse until the repulsive core stops the 
further squeezing of the system. Note, it is impossible to 
represent the denominator as a power series in $R\chi_{0}$
through approximating the expansion by a finite number of terms. This
result is quite obvious since $\rho_{c1}$ is a branch point in the
function $E(\rho )$, which makes it impossible to expand that function
in the vicinity of this point. Therefore, one should try to satisfy
Eq. (6) in order to get proper results for the ground state
calculations in the vicinity of the instability points \cite{sh,as,ksk}. 

We would like to demonstrate here the essential qualitative 
features of atoms in a trap, which can be observed 
experimentally at the densities near the
critical density, $\rho\leq \rho_{c1}$.
It follows from Eq. (9), at least for small $q$ and
$\rho\to\rho_{c1}$, that the susceptibility $\chi(q\to 0,0,1)$, being
proportional to the incompressibility (see Eq. (6)), is divergent.
Therefore, the denominator in Eq. (9) can be expressed in the form
\begin{equation}
1-R(q,0,1)\chi_{0}(q,0)\simeq \alpha_0(\rho/\rho_{c1})+\alpha_1
q^2/p_F^2. \end{equation}
Here $\alpha_0(\rho/\rho_{c1})$ is a positive function such that
$\alpha_0(1)=0$, and $\alpha_1$ is a positive constant.
As a result, the linear response function takes the form,
\begin{equation}
\chi(q\to 0,0,1)|_{\rho\to\rho_{c1}}\propto\frac{1}{q^{2}}.
\end{equation}
Since the linear response function is the density-density correlation
function, we obtain from Eq. (13) that the radius of the
correlation tends to infinity as it must in the 
vicinity of the phase transition point \cite{lanl}. 
For instance, the density
fluctuations $\delta\rho(q)$ induced by an external potential,
$\delta\rho(q)=\chi(q,0)v_{ext}(q)\propto v_{ext}(q)/q^2$.
Thus, an external field $v_{ext}(q)$ even it is of a short range
cannot be screened and produces the fluctuations of the density
of extremely long range. It seems that such a behavior can be
realized in a trap through the observation 
of a strong light scattering by 
the density fluctuations of multi-fermion 
system when $\rho\to\rho_{c1}$. On the other hand, 
the same picture can be observed if the 
density is fixed at some $\rho_0$ value, while the critical density
is driven to this density by adjusting the scattering length
by an external field. If the bare potential were purely 
attractive, the interval of the densities $[0,\rho_{c1}]$
within which the system is stable, would vanish with the growth of $|a|$
since $\rho_{c1}\sim |a|^{-3}$. As a result, in the limit
$a=-\infty$ the density $\rho_{c1}\to 0$, and the considered system
would be completely unstable at any density. As soon as the
scattering length deviates from its infinite value, that is
$+\infty>a>-\infty$ the system comes back to its stable state at list
in the range of the density values $\rho<\rho_{c1}\sim |a|^{-3}$. 
Note, as it follows from our consideration, that any Fermi system
possesses an equilibrium density and energy if the bare
particle-particle interaction contains a repulsive core and its
attractive part is strong enough, so that $a\rightarrow-\infty$.
Indeed, at sufficiently small densities the energy is negative. 
The system collapses (since at these densities the
incompressibility $K\leq 0)$ until the core stops the density growth.
Therefore, the minimal value of the ground state energy must be
negative when the repulsive core switches on to prevent the
system from further collapse. 

Note that trapped alkali Fermi gases are spin polarized and 
cannot interact in s-wave channel. Consider a two component
system realized by mixtures of two Fermi alkali gases,
say $^6$Li and $^{40}$K. We assume that the corresponding  number densities
$\rho_1$ and $\rho_2$ of these gases coincide, $\rho_1=\rho_2=\rho$.
In that case, we come back to
our case: at each level there are two spin-polarized atoms 
interacting in s-wave and with "effective" mass $M^*=M_1M_2/(M_1+M_2)$.
Here $M_1$ and $M_2$ are the particle masses of the alkali
gases under consideration.  As a result upon using Eq. (4),
we obtain $\rho_{c1}\sim |a_c|^{-3}$, where $a_c$ is the scattering
length related to the interaction between the different alkali atoms.
It is possible to consider a large variety of two component Fermi systems
with $\rho_1\neq \rho_2$, or even two component systems composed
of Fermi and Bose gases which can retain the significant features of
the considered one component Fermi system. 
A detailed investigation of such systems
is in progress and will be published elsewhere.
It is worth remarking, that
superfluid correlation cannot stop the system from squeezing, since their
contribution to the ground state energy is negative. After all, the
superfluid correlation can be considered as additional degrees of
freedom which can therefore only decrease the energy. Recent considerations
show that the compressibility related to the superfluid correlation
is positive (see, e.g. \cite{chio} and references therein). Therefore
we do not expect that the contribution coming from the superfluid
correlation can lead to a significant change in our estimate of the
critical density, $\rho_{c1}\sim |a|^{-3}$. While deriving the 
exact relationship, one
has to include a proper treatment of the pairing.

A liquid similar to the model considered above exists in
Nature, viz. liquid $^{3}$He. If a $^{3}$He$_{2}$ dimer exists,
obviously its bounding energy does not 
exceed the bounding energy of a
$^{4}$He$_{2}$ dimer which is $10^{-4}$ meV \cite{kms}. The ground
state energy of liquid helium is about $ 2\ast 10^{-1}$ meV per atom.
Because of this huge difference in binding energies, it is evident
that the contribution coming from the binding
energy of the dimer to the ground state energy of 
the liquid is insignificant. In
fact, numerical calculations show that the pair potential is
rather weak to produce the dimer $^{3}$He$_{2}$ \cite{elg}. Thus, one
can reliably consider an infinite homogeneous system of Helium atoms
as consisting of particles interacting via pair potentials,
characterized by a very large but finite scattering length $|a|\gg
r_{0}$. The following additional remark is appropriate. It seems
quite probable that the neutron-neutron scattering length ($a\simeq
-20$ fm) is sufficiently large to permit the neutron matter to have
an equilibrium energy and density \cite{ksh}. 

In summary, a system of fermions interacting by an artificial
potential has been considered. The qualitative consideration presented
above gives strong evidence that the system becomes unstable at 
densities $\rho\geq \rho_{c1}$. In the vicinity of 
$\rho_{c1}\sim |a|^{-3}$,
there are long-ranged fluctuations of the density which can be
realized in a trap through the strong scattering of light. Our results
suggest that the equation of state of a low density neutron matter
has peculiarities.

Acknowledgments. The visit of VRS to Clark Atlanta University has been
supported by NSF through a grant to CTSPS. AZM is supported by
US DOE, Division of Chemical Sciences, Office of Basic Energy Sciences,
Office of Energy Research.

\end{document}